%% file: vr02Nov_arxiv_replacement.tex
\documentclass[aps,pra,reprint,10pt,a4paper]{revtex4-2}
\usepackage[utf8]{inputenc}
\usepackage[sc,osf]{mathpazo}
\usepackage[T1]{fontenc}
\usepackage{amsfonts}
\usepackage{amssymb}
\usepackage{amsmath}
\usepackage{amsthm}
\usepackage{graphics}
\usepackage{graphicx}
\usepackage{braket}
\usepackage[colorlinks=true,linkcolor=blue,urlcolor=blue,citecolor=blue]{hyperref}

\usepackage{xcolor}
\usepackage{graphicx}
\usepackage{bm}
\usepackage{makeidx} 
\usepackage{amsmath}
\usepackage{amssymb}
\usepackage{underscore}
\usepackage{color}
\usepackage{amsthm}
\usepackage{relsize}
\usepackage{amssymb}
\usepackage{amsfonts}
\usepackage{amsmath}
\usepackage{wasysym}
\usepackage{hyperref}
\usepackage{cleveref}
\usepackage{diagbox}
\hypersetup{colorlinks, citecolor=magenta, filecolor=blue, linkcolor=blue, urlcolor=green}
\usepackage[normalem]{ulem}
\usepackage{float}
\usepackage{multirow}
\newcolumntype{P}[1]{>{\centering\arraybackslash}p{#1}}

\begin{document}

\title{Distribution of entanglement  with variable  range  interactions}

\author{Leela Ganesh Chandra Lakkaraju, Srijon Ghosh, Saptarshi Roy, Aditi Sen (De)}

\affiliation{Harish-Chandra Research Institute, HBNI, Chhatnag Road, Jhunsi, Allahabad 211 019, India}

\begin{abstract}

Distribution of quantum entanglement is investigated for an anisotropic quantum XY model with variable range interactions and in the presence of a uniform transverse magnetic field. We report the possibility of  \emph{qualitative} growth in entanglement between distant sites with an increase in the range of interactions that vary either exponentially or polynomially as the distance between the sites increases. Interestingly, we find that such entanglement enhancement is not ubiquitous and is dependent on the factorization points, a specific set of system parameters where the zero-temperature state of the system is fully separable. In particular, we observe that at zero-temperature, when the system parameters are chosen beyond the pair of factorization points, the increments in entanglement length due to variable range interactions are more pronounced compared to the situation when the parameters lie in between the factorization points.  By employing  the sum of all the bipartite entanglements with respect to a single site, we also show that the shareability of the bipartite entanglements are constrained, thereby establishing their monogamous nature.  Furthermore, we note that the factorization points get reallocated depending on the laws of interaction fall-offs and provide an ansatz for the same. We reveal that the temperature at which the canonical equilibrium state becomes entangled from an unentangled one increases with the increase in the range of interactions, thereby demonstrating enhanced robustness in entanglement against temperature in the presence of long-range interactions and only  when the system parameters are chosen between the pair of factorization points. We apply an energy-based entanglement witness to provide a justification to the observed robustness with temperature.
\end{abstract}
\maketitle
\section{Introduction}
\label{sec:intro}
Towards the end of the last century, it was realized that understanding of quantum mechanics from the  perspective of information theory  is crucial in building quantum technologies \cite{qr1, qr2, nielsenchuang}. 
 It turns out that different forms of non-classicalities \cite{bell, ent1} offered by quantum theory can be useful resources \cite{rt1}, since they can be employed  to achieve higher efficiencies in  certain tasks than their classical  analogs \cite{tele1, dc}. Among all the resource theories developed overtimes, the  theory of quantum entanglement \cite{rt-ent1, rt-ent2, rt-ent3} is  the most prominent one.  Several pioneering protocols like quantum teleportation \cite{tele1, tele2}, quantum dense coding \cite{dc, dcrev}, entanglement-based quantum cryptography \cite{crypto1, crypto2},  one-way quantum computation \cite{oneway1}   were designed by using this novel resource.

Gaining experimental control at the quantum level for scalable implementation of these schemes is one of the major challenges over the last few years. Potential physical systems that lend themselves for such applications include photons \cite{photon-pan}, 
superconducting qubits \cite{supcond}, 
neutral cold atoms  in optical lattices \cite{coldatom1, coldatom4}, ion traps \cite{coldatom5, nmr4}, and nuclear magnetic resonances \cite{nmr1}. 
On the other hand, using many of these revolutionary platforms, quantum spin models  which offer a solid bedrock for achieving  quantum information processing tasks \cite{fazio-rev, ultracold-review}  like quantum state transfer \cite{qst1},  measurement-based quantum computation \cite{mbqc2},  can be realized   with microscopic control over interaction strengths and other system parameters in  laboratories.

Apart from the technological perspective, there are also  fundamental reasons to study quantum spin models by using information-theoretic quantities.  Notably, it was shown that  the nearest neighbor entanglement can serve as the detector of  quantum phase transitions (QPT)\cite{qpt, qptbook2}.
Furthermore, it was found that  for the quantum  spin-$1$ model proposed by Affleck, Lieb, Kennedy, and Tasaki, the AKLT model \cite{aklt}, the entanglement length diverges at the quantum criticality \cite{qpt2}, although the classical correlation length remains finite,  thereby failing to detect the transition. 
Therefore, an analysis of the entanglement profile of quantum spin models is of utmost importance from the dual perspectives of addressing fundamental issues and manufacturing quantum technologies. In the theoretical frontier, several investigations have  been  carried out  \cite{qimeetsqm}, ranging from the thermal behavior \cite{sugoto-vlatko, fazio-rev},  out of equilibrium dynamics \cite{neq}, effects of environmental noise \cite{noise1, noise2}, to name a few. However, most of these studies (cf. \cite{frus-ger, prx1,prb1,prb2,pra1}) are concentrated in two limiting cases, namely models with the nearest neighbor  or  with long-range interactions.

In this paper, we  focus  on a quantum XY spin model  with variable  range interactions,  thereby
 sweeping the entire spectrum of interaction-ranges, starting from the nearest neighbor case to the long-ranged ones, and characterize the distribution patterns of nonlocal resources in terms of entanglement shared between different sites of these models. 
For models with interaction ranges longer than the nearest neighbor case, we consider the subsequent interaction strengths to decrease either exponentially or polynomially (power-law) from the nearest neighbor value with increasing distance between the spins. 
  Each of the two  distributions of relative interaction strengths leads to a set of Hamiltonians for carrying out the investigations. For a given range of interaction, the  profiles of entanglement between different sites are computed when the system is either at zero or at finite temperatures. 
We know that at zero-temperature, the nearest neighbor quantum XY model with a transverse magnetic field  undergoes a quantum phase transition and at the same time, there exists another pair of magnetic field values at   which the zero-temperature states are doubly degenerate and unentangled, known as  factorization points. Note that the existence of factorization points is
 argued to be linked to an \emph{entanglement phase transition}, having no parallel notions in the classical domain \cite{fac-ger}.

We report here that such  factorization points exist even in the presence of  variable range  interactions and get shifted according to the law of decay of the relative  interaction strengths between the sites. Specifically, the gap between the pair of factorization points  increases with the increase in the range of  interactions. 
Interestingly, we observe that factorization points create  two distinct regions in the parameter space according to the spread of entanglement  both in the zero-temperature and the canonical equilibrium states. In the zero-temperature case,  we show that between the pair of factorization points, a longer range of interactions has to be introduced to generate  entanglement between different spins compared to the case beyond the factorization points, irrespective of laws of decays in the interaction strength.  Quantitatively, entanglement lengths also  confirm  distinct features of these two regimes divided via factorization points. 
From a different perspective, we also  investigate the constraints on the shareability of  bipartite correlations by examining the sum of all bipartite entanglements with the first party. We report a non-trivial bound to this quantity that is substantially lower than the algebraic maximum indicating the distribution of entanglement in the zero-temperature state of this model to be monogamous. 
Although factorized states are unique characteristics of the zero-temperature state, we observe here a counter-intuitive consequence of these points in the thermal state. Specifically, we witness that  when the system parameters are chosen between the pair of factorization points, the temperature at which nearest neighbor entanglement  becomes non vanishing, increases with the increase in the range of interactions, thereby revealing enhanced robustness of entanglement for models possessing longer-ranged interactions.
 Surprisingly, such variation of robustness in  the canonical equilibrium state with respect to range is absent when we choose parameters beyond the factorization point. In particular, far from these points, the nearest neighbor states generated via  a variable range of  interactions  becomes entangled at the same temperature, irrespective of the fall-off rates and the other parameters involved in the system. We attempt to explain this improved robustness via a energy-based witness of entanglement \cite{facpt}.

The paper is organized as follows. After a brief discussion of the prerequisites in Sec. \ref{sec:stage}, we move on to study the effects of increasing the interaction range on the trends of entanglement in the zero-temperature  state in Sec. \ref{sec:groundstate}.  Sec. \ref{sec:robust} reports  the increased robustness of entanglement  with the increase in the range of interactions. We draw conclusions in Sec. \ref{sec:conclusion}.

\section{Setting the stage}
\label{sec:stage}
In this section, we describe the model considered in this paper for analysis. Its general properties are explored with a brief characterization of the phases at zero-temperature. We discuss how we tune the range of  interactions, and the fall-off of the relative interaction strengths as the distance between the interacting spins increases. We also talk about other prerequisites required to describe the results of our manuscript. 
In particular, we specify the measure used for quantifying entanglement, and  define also the entanglement length,   the distance upto which entanglement remains finite. The concept of the factorization points is also introduced. 


\begin{figure}[ht] 
\includegraphics[width=\linewidth]{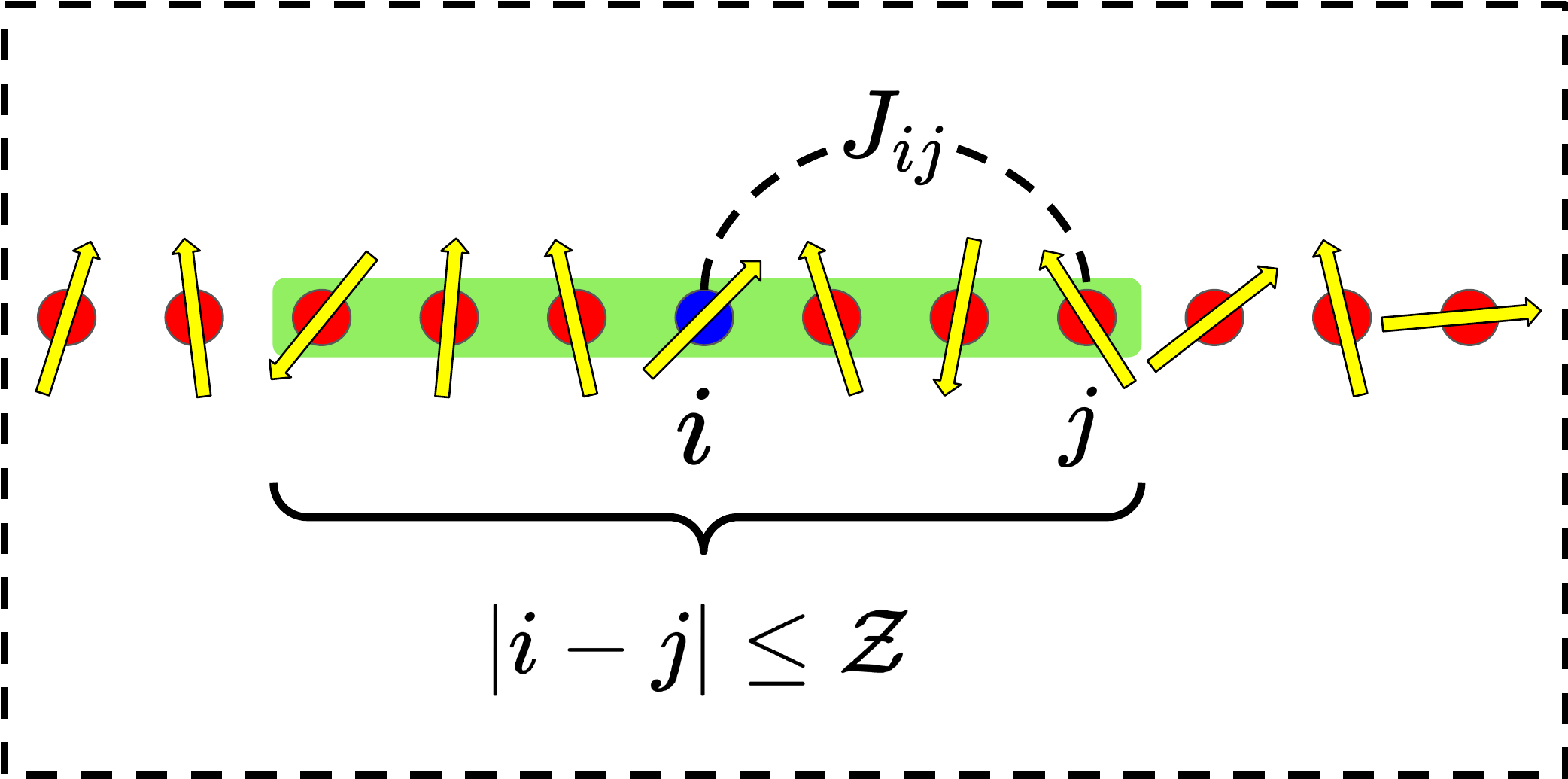}
\caption{ (Color online.) Schematic diagram of a spin model with variable range interactions. 
A  particular case of a variable range model  is displayed where  any given spin, denoted by $i $, interacts with
 $\mathcal{Z} = 3$ nearest  neighbors, as  indicated by the green shaded region $(|i-j| \leq 3)$ about $i$. 
The range of the model can be tuned by varying $\mathcal{Z}$. For the exact form of $J_{ij}$s considered in the manuscript, 
see Eq. \eqref{eq:falloffgen}.}
\label{fig:schematic}
\end{figure}

\subsection{Spin Model with variable-range interactions}
\label{sec:hamiltonian}
Let us consider an anisotropic quantum $XY$ model with variable range interactions 
having periodic boundary conditions described by the following Hamiltonian:
\begin{align}
H = \sum_{\substack{i < j\\|i-j| \leq \mathcal{Z}}}^{N} J_{ij} \Big[ \frac{1+\gamma}{4}\sigma_i^x\sigma_j^x + \frac{1-\gamma}{4}\sigma_i^y\sigma_j^y \Big] + \sum_{i=1}^N \frac{h}{2} \sigma_i^z,
\label{eq:ham1} 
\end{align}
where $\sigma_k^{\hat{n}}$ is the Pauli spin operator associated with the $k^{\text{th}}$ site in the $\hat{n}$ direction with $\sigma_{N+k}^{\hat{n}} = \sigma_k^{\hat{n}}$, and $h$ denotes strength of the uniform magnetic field in the transverse direction. The interaction strength between the $i^{\text{th}}$ and the $j^{\text{th}}$ spin is indicated by $J_{ij}$, while $\gamma$ is the anisotropy parameter which marks the asymmetry in the interaction strengths in  $x$ and $y$ directions. 
The number of sites in the lattice is $N$, and $\mathcal{Z}$ denotes the number of nearest neighbors to which a particular spin couples, which is simply the coordination number.
Thus, for a fixed $\mathcal{Z}$, a given spin $i$ interacts with $\mathcal{Z}$ adjacent spins falling in the region $|i-j| \leq\mathcal{Z}$, see Fig. \ref{fig:schematic}. Therefore, the range of interaction can be varied by changing $\mathcal{Z}$.

In terms of the spin raising and lowering operators $\sigma^{\pm} = (\sigma_x \pm i\sigma_y)/2$, the Hamiltonian in Eq. \eqref{eq:ham1} can be written as
\begin{align}
H=\sum_{\substack{i < j\\|i-j| \leq \mathcal{Z}}}^{N} J_{ij} \Big[ \sigma_i^+\sigma_j^- &+ \sigma_i^-\sigma_j^+ + \gamma(\sigma_i^+\sigma_j^+ + \sigma_i^-\sigma_j^-) \Big] \nonumber \\
 &+ \sum_{i=1}^N h (\sigma_i^+\sigma_i^- - 1/2).
\label{eq:hampar} 
\end{align}
Note that the terms  in the Hamiltonian either counts the number of up (down) spins or flips two spins at a time. Therefore, under this Hamiltonian, the number of up (down) spins modulo $2$ always remain a constant. In other words, it preserves the parity. It makes the Hamiltonian block diagonal into even and odd parity sectors, $H = H^{even} \oplus  H^{odd}$. If the ground state is non-degenerate, it comes from either of the two parity sectors. However, in the case with the ground state being degenerate,  the  eigenstates having minimum energy from both the sectors have the same energy, and hence the ground state has its support from both the parity sectors. Interestingly, our model shows both these kind of features for different ranges of system parameters which we will discuss in subsequent sections.

Let us now consider the relative strength of interactions as the distance between the spins increases. Motivated by the experimental setup, we reasonably assume that the relative interaction strengths decreases with the increase in the distance between the concerned spins $i$ and $j$. Specifically, we consider two qualitatively different fall-off behaviors, given by 
\begin{itemize}
\item the exponential decay: $J_{ij} \sim \alpha_e^{-(|i-j|-1)}$,
\item the power-law decay: $J_{ij} \sim |i-j|^{-\alpha_p}$,
\end{itemize}
where $\alpha_{e (p)}$ denotes the fall-off rates for the exponential and power-law decays respectively. Ultimately, putting everything together, for a given spin $i$, the behavior of interaction strengths $J_{ij}$ depending on the choice of $\mathcal{Z}$ and the decay pattern of relative interaction strengths can be summarized as
\begin{align}
\frac{J_{ij}}{J} = \left\{
 \begin{array}{cc}
 \alpha_e^{-(|i-j|-1)} \text{ or } |i-j|^{-\alpha_p}, &\text{ for } |i-j| \leq \mathcal{Z}  \\
 0, & \text{ otherwise }
\end{array}\right.,
\label{eq:falloffgen}
\end{align}
where $J$ is a constant which corresponds to a ferromagnetic model for $J<0$, the case considered in this manuscript.

In the case of $\mathcal{Z}=1$, the Hamiltonian in Eq. \eqref{eq:ham1} reduces to the well known nearest neighbor anisotropic quantum $XY$ chain, which can be solved analytically \cite{bm1, bm2} for all $N$ and also in the thermodynamic $(N \rightarrow \infty)$ limit. It displays magnetically ordered and paramagnetic phases with a quantum critical point at $\lambda = \pm 1$, where $\lambda = h/J$, a notation we use throughout the manuscript.  
However, the model in Eq. \eqref{eq:ham1}, is in general, not exactly solvable for any other value of $\mathcal{Z}$. Hence for our analysis, we use numerical techniques,  in particular, the Lanczos algorithm \cite{Lanczos}, for finite sized spin chains.
 It employs the idea of Krylov subspaces to tridiagonalize the Hamiltonian 
matrix. By using this method, we can  find a few low-lying eigenstate of the model accurately.and hence can construct the approximate canonical equilibrium state of the model.  The zero-temperature state  of the model \cite{qpt}  is obtained from the canonical state by taking  $\beta \rightarrow \infty$ limit as
\begin{align}
\varrho^0 = \lim_{\beta \rightarrow \infty}\frac{e^{-\beta H}}{\text{tr} (~e^{-\beta H})},
\label{eq:rho}
\end{align}
 where $\beta = 1/k_B T$ is the inverse temperature with $k_B$ being the Boltzmann's constant. 
Notice that the ground state of the model is degenerate for a certain range of $\lambda$ which is yet another reason to use exact diagonalization  method as opposed to other numerical procedures like density matrix renormalization group method  for obtaining  the ground state.  $\varrho^0$ is a \(N\)-party mixed state, containing equal mixtures of all the degenerate ground states of \(H\). When the ground state is non-degenerate, $\varrho^0$ exactly represents the ground state.
To distinguish the former from the symmetry broken ground state, it is usually referred to as the zero-temperature state or the thermal ground state \cite{qpt}, a term which possibly originated from its definition, as given in Eq. \eqref{eq:rho}.



\begin{figure*}[ht] 
\includegraphics[width=\linewidth]{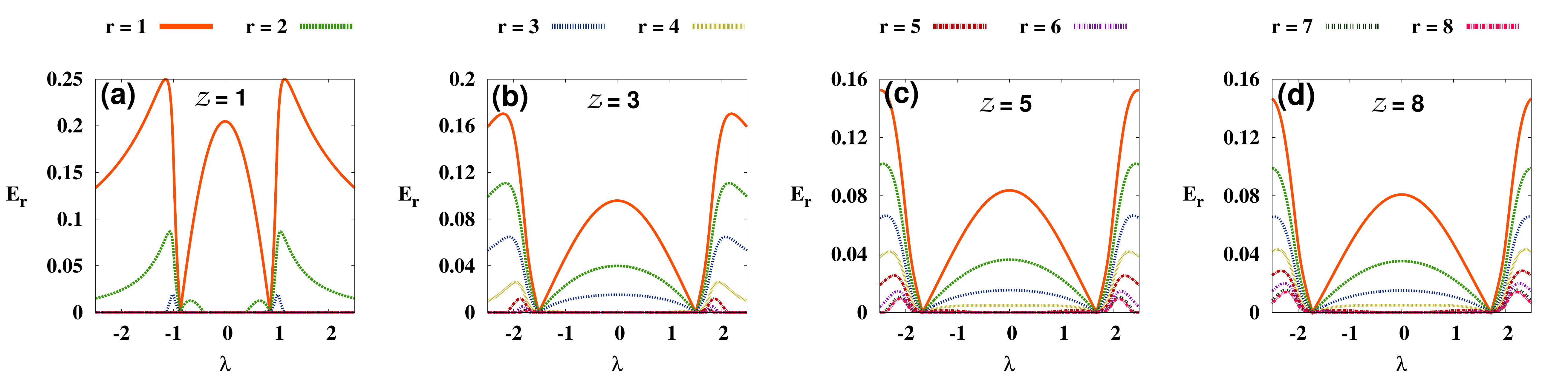}
\caption{(Color online.) Patterns of entanglement distribution in case of  exponential decay of the relative interaction strengths with $\alpha_e = 2$. 
\(E_r, \, r=1 \ldots 8, \) (vertical axis) versus $\lambda $ (horizontal axis).  (a)-(d): For various ranges of interaction,  $\mathcal{Z} = 1, 3, 5,$ and $8$ respectively. Here $N=16$ and \(\gamma =0.5\).  
 Both the axes are dimensionless.}
\label{fig:fig1}
\end{figure*}
\begin{figure*}[ht] 
\includegraphics[width=\linewidth]{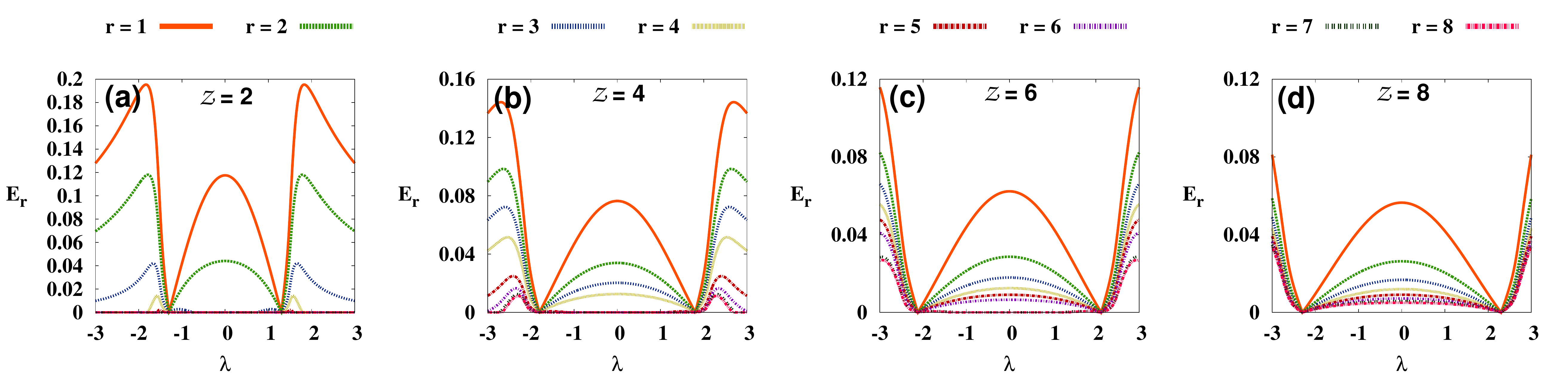}
\caption{(Color online.)  Spread of entanglement of the power-law fall-off  with $\alpha_p = 1$. 
\(E_r \)s (ordinate) are plotted by varying \(\lambda\)  (abscissa). (a)-(d):  $\mathcal{Z} = 2, 4, 6,$ and $8$. All other specifications are same as in Fig. \ref{fig:fig1}. All axes are dimensionless.}
\label{fig:fig2}
\end{figure*}

\section{Consequence of variable range of the zero-\\
temperature state}
\label{sec:groundstate}

We investigate the entanglement profile of the zero-temperature state (thermal ground state) by tuning the strengths of the interaction ranging from the nearest neighbor to the long-ranged one. As mentioned earlier, since the chosen model $(\mathcal{Z} \neq 1)$ cannot be solved analytically, we perform the entire analysis for a finite sized system upto $N = 16$. 
We demonstrate the results  for $N=16$ unless mentioned otherwise, in which the relevant $E_{r}$s, where $E$ is the measure of entanglement, precisely, logarithimic negativity (for further details, see supplementary material, \cite{supple}) are upto  $r=[\frac{N}{2}]$ due to periodicity, 
where $r$ is the distance upto which the entanglement of the reduced density matrix ($\varrho_{r}$) sustains. We examine the behavior of entanglements in two paradigmatic models where the relative interaction strengths follow -- $A.$ the exponential fall-off  for which we present results for $\alpha_e = 2$  for demonstration;  $B.$ the polynomial decay for which we report  the well known Coulomb-type fall-off, i.e., for $\alpha_p = 1$. The overall entanglement profile remains qualitatively similar for other fall-off rates and other finite systems, $N ~(\leq 16)$. We also highlight here that entanglement length gets enhanced and the factorization points are shifted due to the introduction of the variable range interactions. 
We will also discuss the dependence of entanglement on $\alpha_{p}$ as well as $\alpha_{e}$ and the anisotropy parameter $\gamma$.
Since the entire section is devoted to the zero temperature case, we drop the subscript in $\varrho^0$ and  just call it as $\varrho$. 
\subsection{Entanglement profile}
\label{subsec:entpro}

To discuss the consequence of variable range interactions on the entanglement properties of $\{ \varrho_r \}_{r=1}^{\frac{N}{2}}$ at zero temperature, we first consider the case when the relative interaction strengths shows an exponential fall-off. Our aim is to explore the situation  where  for a given  $\mathcal{Z}$, an unentangled  state, $\varrho_r$,   becomes entangled on increasing the interaction range. We find that such a possibility  indeed exists, and call this feature as \emph{activation} where on increasing $\mathcal{Z}$ to $\mathcal{Z}+k$ $(k \geq 1)$, one or more $E_r$s become non-vanishing from the vanishing value.

\emph{A. Exponential fall-off case.} Let us demonstrate the observations for the quantum $XY$ model with $\gamma = 0.5$. The results summarized  below remain qualitatively same for other anisotropy parameters with slight differences which will be addressed in succeeding sections. 

 $1.$ \emph{Nearest-neighbor model.} For nearest-neighbor interaction, i.e.,  $\mathcal{Z}=1$, from  Eqs. (\ref{eq:ham1}) and (\ref{eq:falloffgen}), it is obvious that the Hamiltonians are identical for both the fall-off features, so in this case, one may omit the $e(p)$ labels, and thus henceforth, we call $\lambda^{e(p)}_{f}(1)$ as $\lambda_{f}(1)$ \cite{supple}. We observe that within the factorization points ($-\lambda_f(1) \leq \lambda \leq \lambda_f(1) $), only $E_1$ and $E_2$ are non vanishing, while all other $E_r$s for $r>2$ vanish \cite{fazio-rev}. Note that 
$\lambda_f(1)$ obtained with $N=16$ is  very close to the analytical  value with $N \rightarrow \infty$  (see Table 1 and Fig. \ref{fig:fig1} (a) for details).  On the other hand, outside the factorization points, i.e., when $\lambda < -\lambda_f(1)$ and $\lambda > \lambda_f(1)$ \cite{phasetransrev1}, we discover that $E_r$s show non-zero values, for $r = 1, 2$ and $3$. In both the cases $E_1 > E_2 > E_3$. Moreover, we notice that $ \ E_1^{\max} - \ E_2^{\max} = 0.159$ while $\ E_2^{\max}   - \ E_3^{\max} = 0.059$, where $E_r^{\text{max}}$ denotes the maximal entanglement of $E_r$ in the entire range of $\lambda$ i.e., when $ \lambda \in (-3, 3)$.

$2.$ \emph{Models with $\mathcal{Z} \leq 3$.} Inside the factorization points (corresponding to a given range      $\pm \lambda^e_f (\mathcal{Z})$), entanglement of only a few $\varrho_{r}$s $(r \leq 3)$ gets activated while outside the factorization points, we observe that for $\mathcal{Z} = 3$, entanglement of  $\varrho_r$s becomes non-vanishing for \(r\leq 6\), see Fig. \ref{fig:fig1} (b).

$3.$ \emph{Quantum $XY$ models having $4 \leq \mathcal{Z} \leq 8$.} Progressive activation of long-ranged entanglements (upto $r = 5$) occurs with the variation of $\lambda$ within the respective factorization points. However, $E_r$, $r \geq 7$ with $\mathcal{Z} = 8$ still remains vanishing, thereby showing the absence of activation in presence of exponential fall-off interactions with \(\gamma =0.5\).  It is important to mention here that to obtain nonvanishing  \(E_r,\, \forall \,r\) even in between factorization points, one has to choose high values of \(\gamma\), i.e., towards the Ising limit.  
With $|\lambda|>\lambda^e_f (\mathcal{Z})$, as in the previous scenario,  all $E_r$s are non-vanishing and we notice that  entanglements with $r \geq 6$ possess higher value in this case than that of the model with $\mathcal{Z} \leq 3$, as depicted in  Figs. \ref{fig:fig1} (c) and (d).


\subsubsection*{Role of monogamy in entanglement distribution}

\begin{figure}[h] 
\includegraphics[scale=0.7]{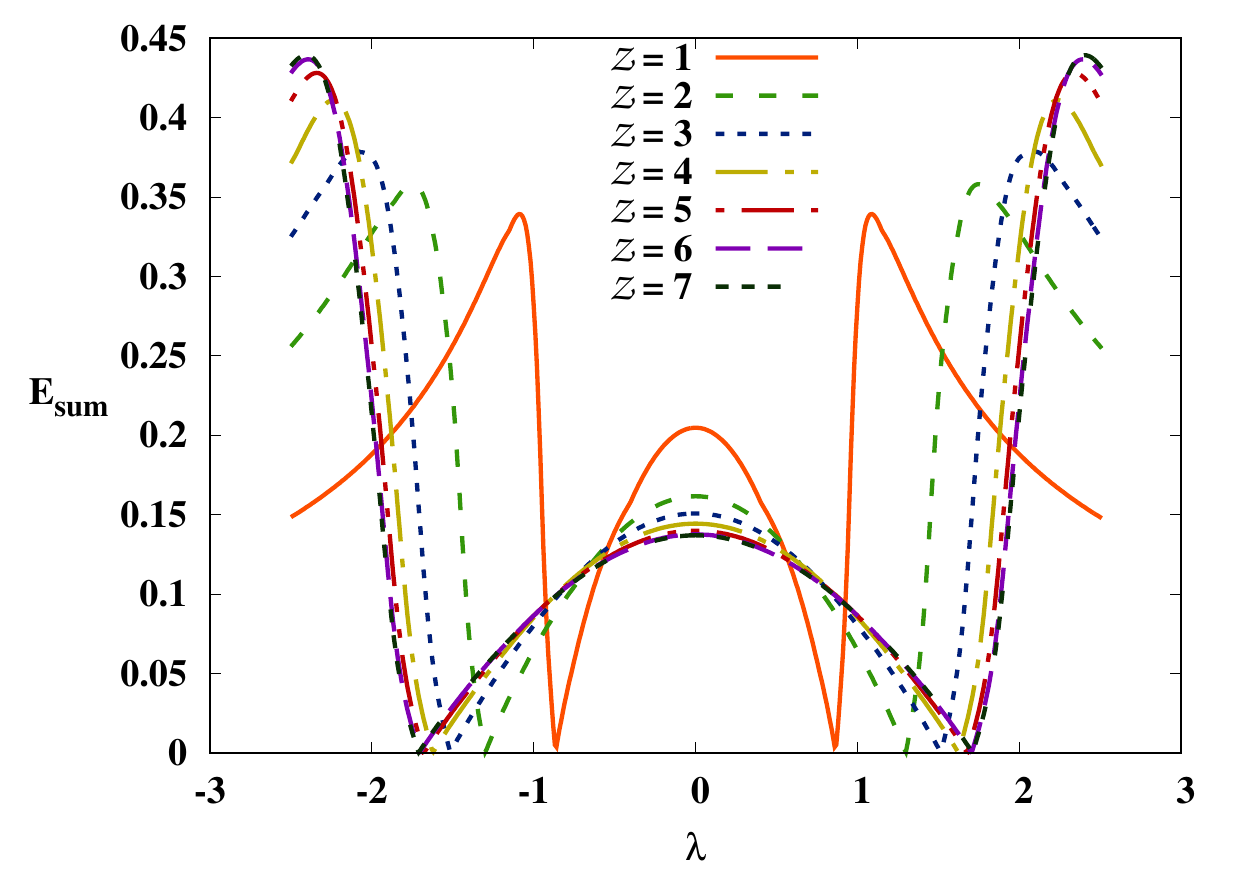}
\caption{ (Color online.)   $E_{sum}$ (vertical axis) with $Z = 1 \ldots 7$ are plotted with respect to $\lambda$ (horizontal axis). Here $N = 14$ with $\gamma = 0.5$. Both the axes are dimensionless.  }
\label{fig:monscore}
\end{figure}
We observe that for all values of $\mathcal{Z}$, the maximal entanglement always comes from the nearest neighbor sector, i.e., $E_1^{\text{max}} > E_{r \geq 2}^{\text{max}}$. Note, however, that $E_1^{\text{max}}$ decreases on increasing the range of interactions, as shown in Fig. \ref{fig:fig1} and
its decrease occurs due to the generation of other long-range entanglement in the model.
 This feature can be  qualitatively explained from the concept of monogamy of entanglement \cite{ckw} which states that a party of a multipartite state cannot share an arbitrary amount of entanglement with other parties. It implies that if a site, say $1$, has a high amount of entanglement shared with another party, say, $2$, party $1$ cannot share a high entanglement content with any other party of a $N$-party state, which clearly justifies the reduction of the nearest neighbor entanglement. 

Quantitatively, it gets reflected in the following way. For any given $\mathcal{Z}$, the value of $\sum_{r=1}^{N-1} E_r= 2 \sum_{r=1}^{N/2} E_r = E_{sum}$ is bounded above by a quantity, $\mathcal{Q}$, substantially smaller than the algebraic maximum of the same,  $N/2 - 1$, i.e.,
\(2\sum_{r=1}^{N/2} E_r \leq  \mathcal{Q}\).
Therefore, when activation of entanglement takes place on increasing $\mathcal{Z}$, more and more $E_r$s start becoming non-zero and hence with some of the $E_r$s which are non-vanishing for small values of $\mathcal{Z}$ has to be  reduced to accommodate the activated $E_r$s, so that the monogamy relation holds.
Note that $\mathcal{Q}$ can itself depend on the the number of sites and the range of the model, see Fig. \ref{fig:monscore} which illustrates the characteristics.
Furthermore, traditionally, $\mathcal{Q}$ is taken to be the entanglement of the zero-temperature state in the $1:$ rest-bipartition, denoted by $E(\varrho_{1:23...N})$. However, in our case,  $\mathcal{Q} >  E(\varrho_{1:23...N})$ except  the region, just outside the factorization point.
Hence, our analysis reveals that the actual bounds  are more complex function of state parameters than the one considered in the traditional monogamy inequality, i.e. $\mathcal{Q} =  E(\varrho_{1:23...N})$, making the situation more interesting.


 \emph{Difference in maximal entanglement.}
The difference between the maximal entanglements of $\varrho_r$ and $\varrho_{r+1}$ is given by 
$\Delta_{E_r} = E_r^{\text{max}} - E_{r+1}^{\text{max}},$ 
for different $\mathcal{Z}$. We observe that for a given $\mathcal{Z}$, $\Delta_{E_r}$ decreases progressively with increasing $r$. Furthermore, for all $r$, $\Delta_{E_r}$, if non-zero, decreases on increasing $\mathcal{Z}$. We believe that such features are seen owing to the comparatively small entanglement values for larger choices of $\mathcal{Z}$ and $r$ as well as monogamy of entanglement, as discussed before. 

\emph{B. Models with power-law fall-off. } Let us now move to the quantum spin models with the power-law decay of relative interaction strengths. We observe qualitatively similar features in the behavior of entanglement as seen in the  exponential ones (see Fig. \ref{fig:fig2}). However, there are some contrasting characteristics like pronounced activation of entanglement  observed due to the slower decay of subsequent interaction strengths in the case of the power-law fall-off compared to the exponential ones. In particular, if one turns on all the interaction terms in the Hamiltonian $(\mathcal{Z}=8)$, a finite amount of entanglement is generated with the variation of $\lambda$ between factorization points even in case of $\varrho_{6}$, $\varrho_{7}$ and $\varrho_{8}$, which is not true for the exponential case (comparing Figs. \ref{fig:fig1} (d) and \ref{fig:fig2} (d)). On the other hand, beyond the factorization points, activation features for both the decay types are almost identical, although the entanglement contents  for the power-law fall-off are comparatively lower than that of the exponential ones. However, the substantial decrease in the difference between $E_r$ and $E_{r+1}$, i.e., $\Delta_{E_r}$ happens in case of power-law decay which is not the case for exponential ones. 

As mentioned before, the behavior of entanglement in these classes of quantum spin models depend on $\lambda$, $\gamma$, $\alpha$ and $\mathcal{Z}$. Upto now, we have discussed the trends of $E_{r}$ with respect to $\lambda$ and $\mathcal{Z}$, by fixing $\gamma$ and $\alpha$. Although the observations remain qualitatively similar for other choices of system parameters, there are some subtle differences that can be seen on changing the anisotropy parameter $\gamma$ and the fall-off rate $\alpha_{e (p)}$. We will analyze these differences in subsequent sections.
Nevertheless, what emerges out of our analysis, and which remains true irrespective of the choice of system parameters and fall-offs  is that the factorization points divide  the magnetic fields  into two regions having qualitatively distinct entanglement profiles. 
Specifically, increase in the range of  interactions outside the factorization points stimulates   entanglement in longer spatial sites much faster than the scenario within the factorization points  for small values of  anisotropy parameters while in presence of high anisotropy, entanglements over longer range can be generated in both the regimes. 

\subsection{Enhanced  entanglement length}

The preceding analysis confirms the activation of entanglement  by introducing the variable range of interactions in a subjective manner. Let us quantify the production of long-range entanglement by computing the entanglement length, $\xi$, which is defined via $E_r = a + be^{-\frac{r}{\xi}},$ with $a$ and $b$ being the constants  \cite{supple}, for different values of $\mathcal{Z}$. 

\begin{figure}[ht] 
\includegraphics[width=0.9\linewidth]{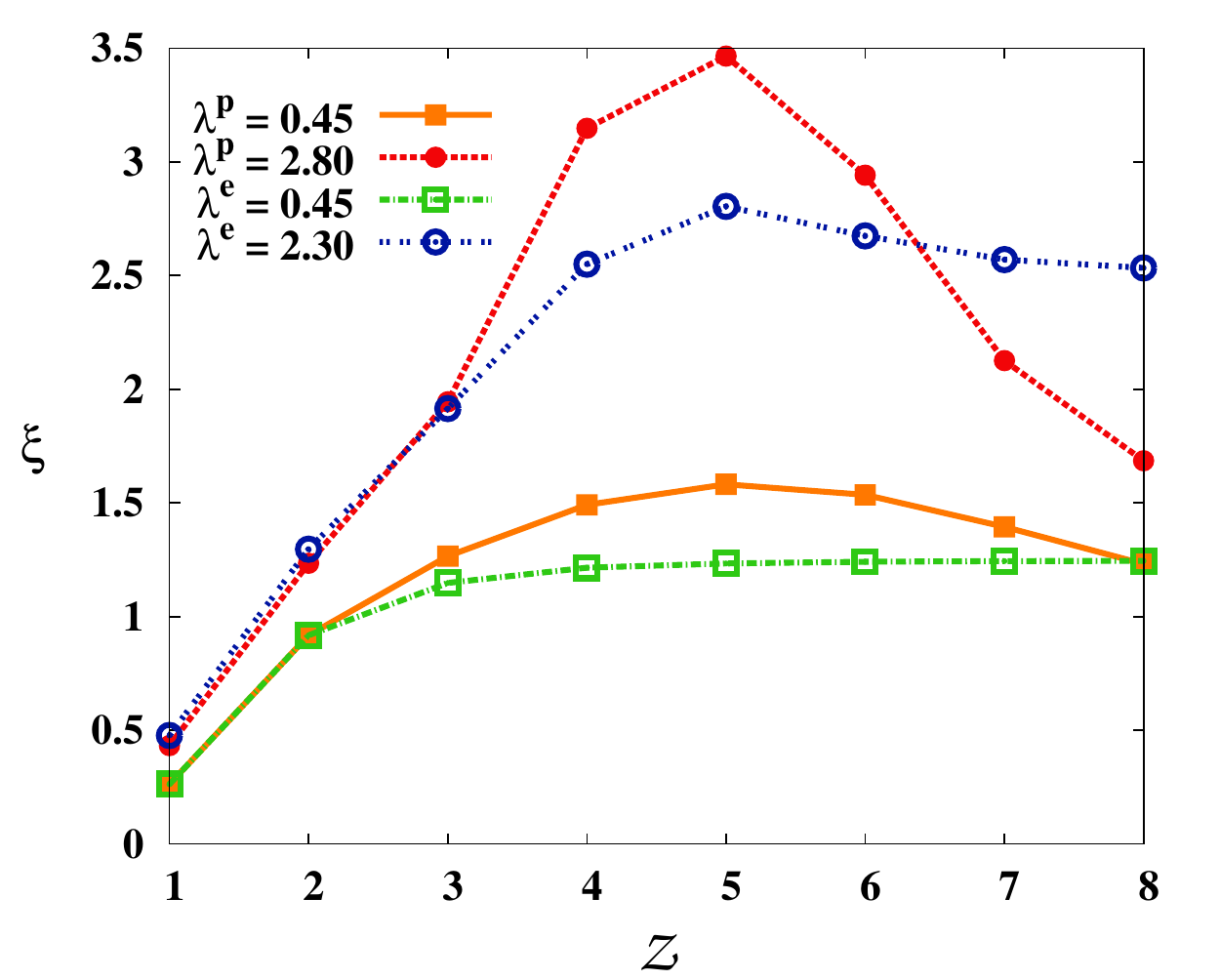}
\caption{(Color online.) Entanglement length, \(\xi\), (in the $y$-axis) with respect to range of interactions, $\mathcal{Z}$ (in the $x$-axis). 
For demonstration, $\lambda^{e}$s and $\lambda^p$s are  chosen within the pair of factorization points and outside of them. Note that for clarity, we mark \(\lambda\)s by superscripts \(e\) and \(p\)  to denote exponential and power-law fall-offs respectively. Other specifications are same 
as in Figs. \ref{fig:fig1} and \ref{fig:fig2}. Both the axes are dimensionless.}
\label{fig:length}
\end{figure}

\emph{Exponential fall-off case.} Since we know that for a fixed  values of $\gamma$, $\mathcal{Z}$ and $\alpha_e$, the activation has a different nature inside the pair of factorization points and beyond,  we examine $\xi$ by setting $\lambda^e =0.45 < |\lambda^e_f(\mathcal{Z})|$ and $\lambda^e = 2.3 > |\lambda^e_f(\mathcal{Z})|$ (To differentiate between the strengths of magnetic fields in  cases of exponential and power law fall-offs, we use superscripts \(e\) and \(p\) in \(\lambda\)s  for referring exponential and power law decays respectively.) When $\lambda^e = 0.45$, $\xi$ increases monotonically with $\mathcal{Z}$ and after $\mathcal{Z} \geq 4$, the increase in \(\xi\) is almost insignificant (the change in the order of $10^{-2}$)  while significantly higher value of $\xi$ is obtained with $\lambda^e = 2.3$ (see Fig. \ref{fig:length}). In the latter case, we also witness nonmonotonic behavior of $\xi$ with $\mathcal{Z}$ and the maximal value of $\xi$ is obtained for $\mathcal{Z} = 5$ with $N = 16$.  Both the situations clearly indicates the production of  entanglement over long distance due to introduction of variable range interactions. Note that although we fix $\lambda^e$ values for illustration,  similar patterns in entanglement length also emerge  for other values of $\lambda^e$, chosen from inside and outside of the factorization points.

%

\emph{Power-law fall-off case.} 
In sharp contrast with the exponential case, when $\lambda^p$ is chosen between two factorization points and beyond, entanglement lengths always show nonmonotonicity with $\mathcal{Z}$. As one can infer from the entanglement profiles, $\xi$ posses higher value with $\mathcal{Z}$ when $\lambda^p$ is outside the pair of factorization points than that of the case chosen inside the factorization points.  Note that in this case, 
\(\lambda^p=2.8> |\lambda^p_f(\mathcal{Z})| \) is chosen, since the factorization points shift according to the laws of fall-off, as will be seen in the next section.  Both the scenarios clearly confirm the spread of entanglement between distant sides due to the  variable range interactions, thereby illustrating the importance of long range interactions in   generation of resources.

\noindent
\textbf{Note:} Upto now, all the analysis are presented for a fixed value of $\alpha_{e(p)}$ and $\gamma$. However, we check the behavior of entanglement pattern for different values of anisotropy parameter as well as fall-off rates and find out that the behavior is qualitatively similar as before. For a detailed discussion, see supplementary material \cite{supple}.

\subsection{Shifts in factorization points}

All the analysis in the preceding section clearly demonstrates that factorization points play a crucial role in the trends of the entanglement distribution \cite{facrev1, facrev2,facrev3} . Let us determine the effect of $\mathcal{Z}$ on the factorization points. 
In the nearest neighbor case,  factorization points are given by $\lambda_f(1) = \pm \sqrt{1-\gamma^2}$. Note that, for any $\gamma \neq 0$, $\lambda_f(1) \leq \lambda=1$, the quantum critical point. Therefore, the factorization points for $\mathcal{Z} = 1$ always lie inside the magnetically ordered phase \cite{facrev4, facrev5}. We will come back to this point later. 
 
\noindent\textbf{Proposition.}

Considering  variable range interactions, the factorization point where the eigenstate with minimum energy is absolutely separable reads as
\begin{equation}
\lambda^{e(p)}_f(\mathcal{Z}) = \pm \sqrt{1-\gamma^2}\sum_{\mid i - j \mid = 1}^{\mathcal{Z}} \frac{J_{ij}}{J},
\label{eq:facptgen}
 \end{equation} 
  for any spin index $i$.

\begin{proof}: 

Consider the arbitrary product state of $N$ qubit spins as
\begin{equation}
|\psi_P\rangle = \mathlarger{\mathlarger{\mathlarger{\otimes}}}_i \cos \frac{\theta_i}{2} |0\rangle +e^{i\phi_i} \sin \frac{\theta_i}{2} |1\rangle.
\end{equation}
Let us find out the inner product of the Hamiltonian with this state, which gives the energy, parameterized by  $\{\theta_i,\phi_i\} $. We then minimize the energy by varying the set  $\{\theta_i,\phi_i\} $, i.e., 
\begin{equation}
E_P = \min_{\{\theta_i,\phi_i\}} \langle\psi_P|H|\psi_P\rangle.
\end{equation}
On the other hand, we can also find that the minimum eigenvalue of the Hamiltonian when  $\lambda = \lambda^{e(p)}_f(\mathcal{Z})$, labelled as $E(\lambda^{e(p)}_f(\mathcal{Z}))$ which matches with \(E_P\), thereby confirming that the zero-temperature state is factorized.  With $N = 14$ and the exponential case of $\lambda^e = 2$,  the results of \(E_P\) and $E(\lambda^{e}_f(\mathcal{Z}))$    are tabulated for given $\mathcal{Z}$ (Table. \ref{facTable}). 

\begin{table}
\begin{center}
\begin{tabular}{|P{0.8cm}|P{2cm}|P{2cm}|P{2cm}|}
\hline
$\mathcal{Z}$ & $E_P$     & $E(\lambda^e_f)$  \\ \hline
1 & -6.999  & -7      \\ \hline
2 & -10.499 & -10.5   \\ \hline
3 & -12.249 & -12.25  \\ \hline
4 & -13.125 & -13.125 \\ \hline
5 & -13.562 & -13.563 \\ \hline
6 & -13.781 & -13.781 \\ \hline
7 & -13.863 & -13.863 \\ \hline
\end{tabular}
\end{center}
\caption{ $E_P$ is the minimum value of inner product of product state with the Hamiltonian and $E(\lambda^e_f)$ is the energy of the minimum eigenvalue at $\lambda = \lambda^e_f $, the predicted factorization point.}
\label{facTable}
\end{table}
\end{proof}

For the exponential fall-off, it reduces to
\begin{equation}
\lambda^e_f(\mathcal{Z}) =\pm \sqrt{1-\gamma^2}\sum_{k = 0}^{\mathcal{Z}-1} \frac{1}{\alpha_{e}^{k}}.
 \end{equation} 
Note that it represents a geometric progression which can be summed easily.
 If we choose $\alpha_e = 2$,
 \begin{align}
 \lambda^e_f(\mathcal{Z}) = \pm 2\big(1-2^{-\mathcal{Z}}\big)\sqrt{1-\gamma^2}.
 \end{align}
 
Let us now compute and compare the predicted shifts in the factorization points, $\lambda^e_f(\mathcal{Z})$, to that obtained, via analysis with finite-size system $N = 16$, for various choices of $\gamma$, denoted by $\Lambda^e_f(\mathcal{Z})$,  see Table. \ref{table:facpt-exp}. We compute  $\Lambda^e_f(\mathcal{Z})$ upto a precision of $\pm 0.01$, which is our chosen step size in the $\lambda$-axis for numerical simulations while we  round off the predicted value, $\lambda^e_f(\mathcal{Z})$ upto the third decimal point. We observe that the predicted and observed factorization values almost exactly coincide upto the third significant digits. Note that on increasing the range of interaction, the gap between two factorization points, $|\lambda| < \lambda^e_f(\mathcal{Z})$ increase. Moreover, by checking the order parameter, $m_x$  \cite{mx, qptbook2}, we confirm that like the case with $\mathcal{Z} = 1$, the pair of factorization points always lie within the magnetically ordered phase for any given range of interactions.\\

\begin{table}[h]
\begin{center}
\begin{tabular}{|P{0.2cm}|P{0.9 cm}|P{0.9 cm}|P{0.9 cm}|P{0.9 cm}|P{0.9 cm}|P{0.9 cm}|}
\hline
\multirow{2}{*}{$\mathcal{Z}$} &  \multicolumn{2}{c}{$\gamma = 0.2$} \vline & \multicolumn{2}{c}{$\gamma = 0.5$} \vline & \multicolumn{2}{c}{$\gamma = 0.8$} \vline \\ \cline{2-7}
     &  $\lambda^e_f$ & $\Lambda^e_f$ & $\lambda^e_f$ & $\Lambda^e_f$ & $\lambda^e_f$ & $\Lambda^e_f$ \\
\hline
1 & 0.980 & 0.98 & 0.866 & 0.86 & 0.600  & 0.60  \\ \hline
2 & 1.470 & 1.47 & 1.299  & 1.30  & 0.900  & 0.90  \\ \hline
3 & 1.715 & 1.71 & 1.516 & 1.51 & 1.050 & 1.05 \\ \hline
4 & 1.837 & 1.84 & 1.624 & 1.62 & 1.125 & 1.12 \\ \hline
5 & 1.898 & 1.90  & 1.678 & 1.69 & 1.163 & 1.16 \\ \hline
6 & 1.929 & 1.93 & 1.705 & 1.70 & 1.181 & 1.18 \\ \hline
7 & 1.944 & 1.94 & 1.719 & 1.72 & 1.191 & 1.19 \\ \hline
8 & 1.952 & 1.95 & 1.725 & 1.72 & 1.195  & 1.19 \\ \hline
\end{tabular}
\end{center}
\caption{Predicted $(\lambda^e_f)$ and observed $(\Lambda^e_f)$ factorization points for three different values of the anisotropy parameters, $\gamma = 0.2, 0.5,$ and $0.8$. $\mathcal{Z} \in [1,8]$ when the relative interaction strengths show an exponential fall-off with $\alpha_e = 2$.}
\label{table:facpt-exp}
\end{table}

In case of power-law decay, the predicted factorization-point formula in Eq. \eqref{eq:facptgen} takes the following form:
\begin{equation}
 \lambda^p_f(\mathcal{Z})  =\pm  \sqrt{1-\gamma^2}\sum_{k = 1}^{\mathcal{Z}} \frac{1}{k^{\alpha_p}}.
 \end{equation} 
Note that unlike the exponential fall-off case, it cannot  be summed for a general $\alpha_p$. For $\alpha_p = 1$, we have
 \begin{equation}
 \lambda^p_f(\mathcal{Z})  = \pm \sqrt{1-\gamma^2}H(\mathcal{Z}),
 \end{equation} 
 where $H(n)$ denotes the $n^{\text{th}}$ Harmonic number, defined as the sum of reciprocals of the first $n$ natural numbers. Like in the exponential case, we make a comparative study of predicted and observed factorization points in Table. \ref{table:facpt-power}. Again, like before, we observe the widening of the gap between the factorization points on increasing $\mathcal{Z}$. Note that we get a very good agreement between the predicted and observed factorization points upto our precision of $\pm 0.01$. 
\begin{table}[h]
\begin{center}
\begin{tabular}{|P{0.2cm}|P{0.9 cm}|P{0.9 cm}|P{0.9 cm}|P{0.9 cm}|P{0.9 cm}|P{0.9 cm}|}
\hline
\multirow{2}{*}{$\mathcal{Z}$} &  \multicolumn{2}{c}{$\gamma = 0.2$} \vline & \multicolumn{2}{c}{$\gamma = 0.5$} \vline & \multicolumn{2}{c}{$\gamma = 0.8$} \vline \\ \cline{2-7}
     &  $\lambda^p_f$ & $\Lambda^p_f$ & $\lambda^p_f$ & $\Lambda^p_f$ & $\lambda^p_f$ & $\Lambda^p_f$ \\
\hline
$1$ & $0.980$ & $0.97$ & $0.866$ & $0.86$ & $0.600$ & $0.60$\\
\hline
$2$ & $1.470$ & $1.47$ & $1.299$ & $1.30$ & $0.900$ & $0.90$\\
\hline
$3$ & $1.796$ & $1.79$ & $1.588$ & $1.59$ & $1.100$ & $1.10$\\
\hline
$4$ & $2.041$ & $2.04$ & $1.804$ & $1.80$ & $1.250$ & $1.25$\\
\hline
$5$ & $2.237$ & $2.24$ & $1.977$ & $1.98$ & $1.370$ & $1.37$\\
\hline
$6$ & $2.400$ & $2.40$ & $2.122$ & $2.12$ & $1.470$ & $1.47$\\
\hline
$7$ & $2.540$ & $2.54$ & $2.245$ & $2.24$ & $1.556$ & $1.56$\\
\hline
$8$ & $2.663$ & $2.60$ & $2.353$ & $2.30$ & $1.631$ & $1.60$\\
\hline
\end{tabular}
\end{center}
\caption{Predicted $(\lambda^p_f)$ and observed $(\Lambda^p_f)$ factorization points for $\gamma = 0.2, 0.5,$ and $0.8$ with  $\mathcal{Z} \in [1,8]$. Here the relative interaction strengths show a power-law fall-off with $\alpha_p = 1$.}
\label{table:facpt-power}
\end{table}

%


\noindent \textbf{Note:} The typical values of entanglement $(\sim 0.25)$ reported in our work are quite generic for the model under consideration. However, to understand the physics of these models, for example, the detection of quantum phase transitions, dynamical quantum phase transition, the value of entanglement is not the quantity of interest, rather the analyticity of entanglement near the transition points becomes the relevant marker. Apart from theoretical insights, these models can be experimentally realized in ultracold atoms trapped in optical lattices as discussed in  Sec. \ref{sec:intro}, thereby making them a potential candidate for quantum technologies.


\section{Robustness of entanglement to temperature}
\label{sec:robust}

The exact zero temperature regime is obviously an idealization that cannot be realized in practice. 
Systems inevitably suffer from thermal noise, thereby, in general, reducing the quantum correlations.  
The state in thermal equilibrium with the bath reads as
$\varrho = \frac{e^{-\beta H}}{\text{tr} (~e^{-\beta H})} = \frac{\sum_{i=1}^{2^N} e^{-\beta e_{i}}|e_{i}\rangle \langle e_{i}|}{\sum_{i=1}^{2^N} e^{-\beta e_{i}}} $,
where $H$ is the model Hamiltonian as defined in Eq. \eqref{eq:ham1}, and $\lbrace e_{i}, |e_{i}\rangle \rbrace$ are the eigenvalues and eigenvectors of $H$. To make \(\beta\) dimensionless, we refer \(\beta/J\) as \(\beta\).   
 The typical investigation in this context is to identify the temperature upto which the thermal state remains entangled \cite{robustness} which measures the robustness of  entanglement in the canonical equilibrium state against thermal fluctuations \cite{thermrev1, thermrev2}. In our case, we also study the role of the interaction range and the choice of system parameters on the observed robustness.
Interestingly, we again report that the factorization points obtained at the zero temperature plays a crucial role in determining the thermal entanglement profile.
\begin{figure}
\includegraphics[width=\linewidth]{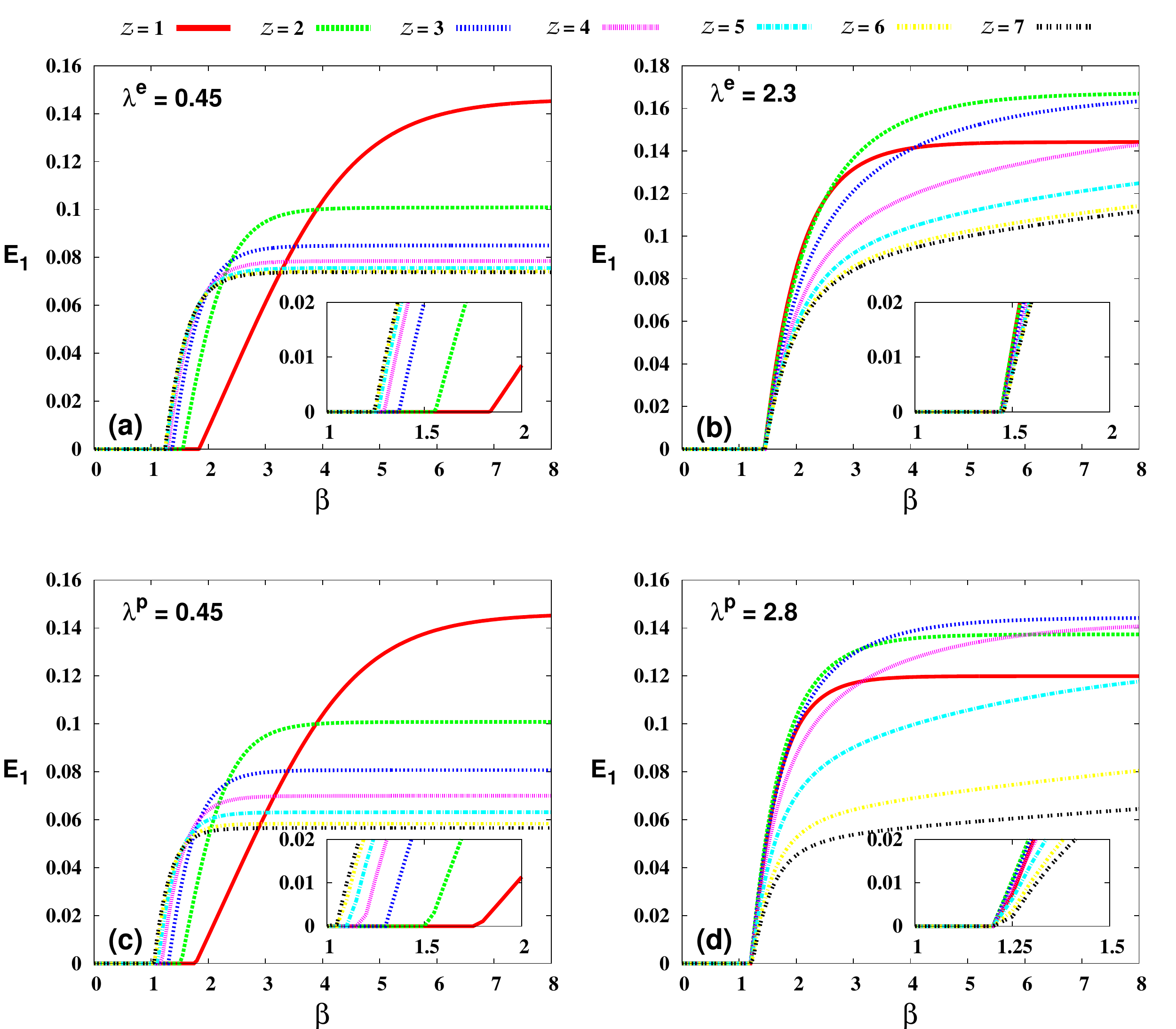}
\caption{ (Color online.)  (a)-(b) Nearest neighbor entanglement, $E_1$ (abscissa) vs. $\beta$ (ordinate) for exponential decay of relative interaction strengths, 
\(\lambda^e =0.45\) and  \(\lambda^e = 2.3\) respectively. 
(c)-(d) \(E_1\) against \(\beta\) with \(\lambda^p =0.45\) and  \(\lambda^p = 2.8\) respectively. Here  $N =14$ and  $\gamma = 0.5$. The choices  of \(\lambda^{e(p)}\)s are same as in  Fig. \ref{fig:length} (for details, see text).  Both the axes are dimensionless.}
\label{fig:thermalexp}
\end{figure}

Specifically, for both exponential and power-law fall-offs, when we choose a $\lambda$ value within the factorization points of the zero-temperature state, we observe an increased robustness of nearest neighbor entanglements obtained for increased ranges of interaction, as shown in  Figs. \ref{fig:thermalexp} (a) and (c). 
To quantify robustness, we introduce a quantity, named as \emph{critical temperature}, $1/k_B \beta^*_{\mathcal{Z}}$, at which the nearest neighbor entanglement, $E_{1}$, for a given $\mathcal{Z}$ starts becomes nonvanishing. We find $\beta^*_{\mathcal{Z}}$ decreases with the increase of $\mathcal{Z}$. 
It implies that with the increase of range of interactions, nearest neighbor entanglement remains nonvanishing even in presence of higher temperature. For example, with $\gamma = 0.5$,  and \(\alpha_e=2\), we find $E_{1} > 0$ when $\beta^{*}_7 = 1.25$ for \(\mathcal{Z} =7\) while  $\beta^*_{3} = 1.37$ for \(\mathcal{Z} =3\).
In stark contrast, when the $\lambda$ is chosen from outside the factorization points, no such robustness is observed and far from factorization points, $E_1$ for all $\mathcal{Z}$ becomes nonzero from the same critical temperature, see Figs. \ref{fig:thermalexp} (b) and (d).
We want to stress here that such a strong dependence of finite temperature physics on the property of the zero-temperature state, namely factorization points, is highly nontrivial.

Note that  results presented here is for $N =14$ for which exact diagonalizations cannot accurately give all the $2^{14}$ eigenvalues and eigenvectors. Therefore, we consider an approximate canonical equilibrium state of the form
 \begin{align}
\varrho \approx  \frac{\sum_{i=1}^{m} e^{-\beta e_{i}}|e_{i}\rangle \langle e_{i}|}{\sum_{i=1}^{m} e^{-\beta e_{i}}} ,
\label{eq:rhoth2}
\end{align}
where $m < 2^{14}$ corresponds to the lowest $m$ eigenvalues of $H$, obtained using the Lanczos algorithm as discussed in Sec. \ref{sec:hamiltonian}. We fix the values of $m$ by examining the results from two different angles, namely convergence and continuity, as follows:
\begin{enumerate}
\item We first track $\beta^*_{\mathcal{Z}}$ by changing $m$ starting from $m = 200$, increasing it in steps of $25$. We claim convergence when even on increasing $m > m'$, $\beta^*_{\mathcal{Z}}$ changes insignificantly, in particular change in $\Delta \beta^*_{\mathcal{Z}} =  |\beta^*_{\mathcal{Z}} (m) -  \beta^*_{\mathcal{Z}}(m')| <10^{-4}$, for all $\mathcal{Z}$.

\item Secondly, we compare the $\beta^*_{\mathcal{Z}}$s obtained with $m=m'$ for $N = 14$  with the values obtained for $N = 8, 10$ and $12$. Note that the results  for $N \leq 12$ are obtained using the exact diagonalization technique and hence $\lbrace e_i, |e_i\rangle \rbrace$ are exactly obtained. The $\beta^*_{\mathcal{Z}}$ values and the qualitative entanglement features with $\beta$ in each of the cases are comparable which assures consistency of our results via continuity.
\end{enumerate}
For both exponential $(\alpha_e = 2)$ and power-law $(\alpha_p = 1)$ decays, $m'$ turns out to be $300$. 


\subsubsection{Entanglement Witness }

\begin{figure}
\includegraphics[width=9cm, height=4.5cm]{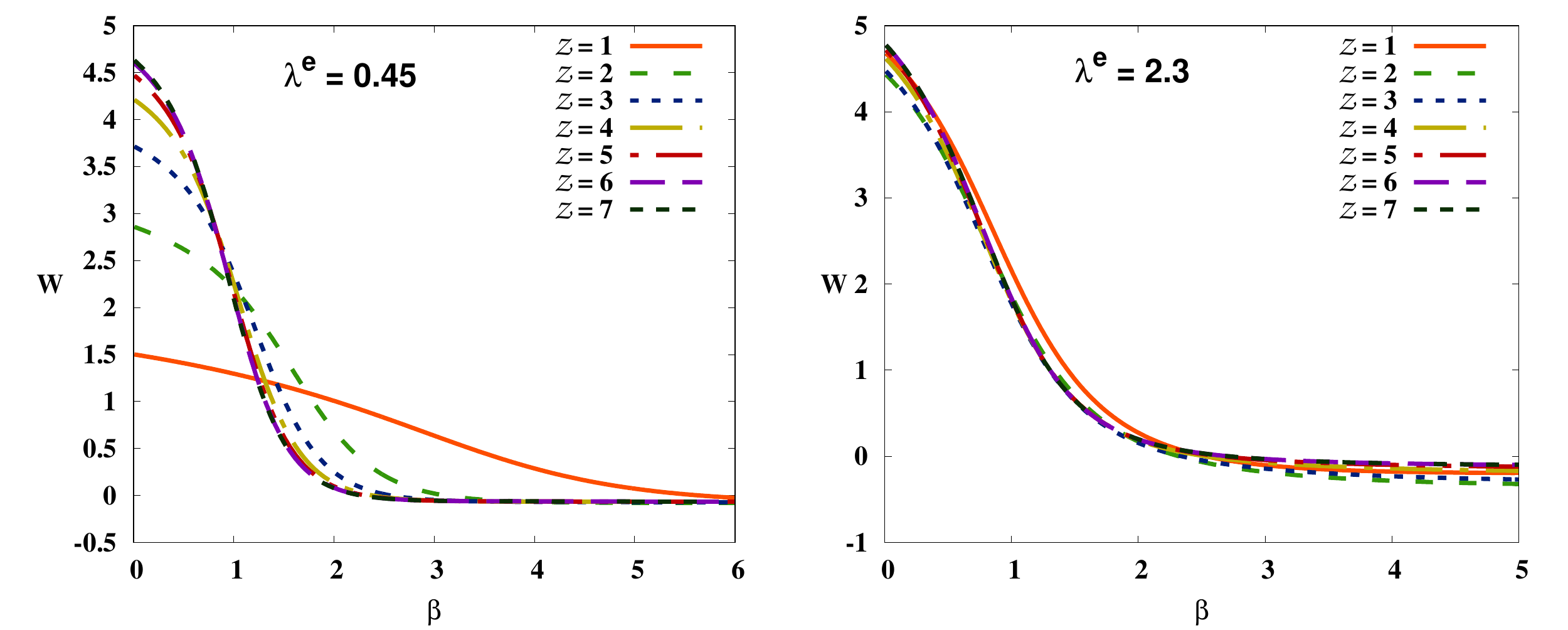}
\caption{ (Color online.) Witness $W$ (abcissa) vs. $\beta$ (ordinate) for exponential decay of relative interaction (Left) \(\lambda^e =0.45\) and   (Right) \(\lambda^e = 2.3\). Here  $N =14$ and  $\gamma = 0.5$. The choices  of \(\lambda^{e}\) are same as in  Fig. \ref{fig:thermalexp} (for details, see text).  Both the axes are dimensionless.}
\label{fig:tempWitness}
\end{figure}

Let us consider a witness operator constructed based on entanglement gap \cite{facpt}.
\begin{equation}
W = \langle H \rangle- \min_{\{\mbox{sep}\}}E_{sep},
\label{eq:witness}
\end{equation}
where $\langle H \rangle$ is the energy with respect to the thermal state, $\varrho$, in Eq.  (\ref{eq:rhoth2}) for a given $\beta$ and $E_{sep}$ is obtained  after minimizing the  energy over the set of fully factorized states for a given configuration of $H$. It was argued that the witness operator is capable to detect entanglement of the global state by giving negative  value for a given $\beta$.  Although this witness operator detects entanglement of the global state,  we observe that its behavior is qualitatively  similar to that of the reduced bipartite states, \(\{E_r\}\) , obtained from the thermal state (see Fig. \ref{fig:thermalexp} for nearest neighbor entanglement profiles with temperature for different \(\mathcal{Z}\)). 
In terms of witness,  when  $|\lambda|<\lambda_f$ and temperature is high, the rate at which $E_{sep}$ increases with $\mathcal{Z}$ is much higher than the  rate at which energy of the thermal state ($\langle H \rangle$) increases with $\mathcal{Z}$,   thereby indicating witness to be negative and showing robustness in entanglement with temperature (see Fig. \ref{fig:tempWitness} (left)). 
On the other hand, 
when  $|\lambda|>\lambda_f$, both the rates are similar, and hence the temperature at which $W<0$ is almost same, thereby confirming the observation that the non-zero entanglement is found at almost the same temperature value irrespective of  $\mathcal{Z}$ as depicted in Fig. \ref{fig:tempWitness} (right)
.

\subsection{Rigidity of robustness to variations of magnetic field}

\begin{figure*}[ht] 
\centering
\includegraphics[width=\linewidth]{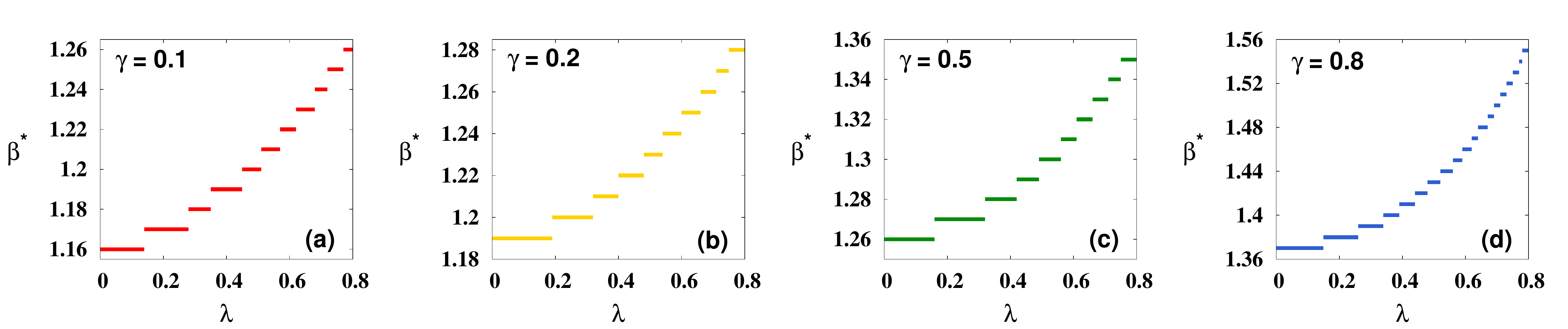}
\caption{(Color online.)  Plateaus representing constant  \(\beta^*\) (ordinate) with respect to magnetic field, \(\lambda\), (abscissa) for different values of the anisotropy parameters. Specifically, $\gamma = 0.1$ in (a), $\gamma = 0.2$ in (b), $\gamma = 0.5$ in (c), and $\gamma = 0.8$ in (d). For all the $\gamma$ values, when $\lambda$ is  close to $0$ offer the highest critical temperature. Similarly, low $\gamma$ values yield comparatively higher critical temperatures than that of the high values of \(\gamma\). The figures are for the exponential fall-off with $\alpha_e =2$, and $N = 12$.   All axes are dimensionless.}
\label{fig:plateau}
\end{figure*}

Let us first recall that the robustness was dependent on the interaction range when $\lambda$ lies within the factorization points of the thermal ground state (see Figs. \ref{fig:thermalexp} (a) and (c)). In particular, longer-ranged interactions offer higher critical temperatures, i.e., enhanced robustness and hence the highest robustness is obtained  for the maximal range, $\mathcal{Z}_{\max}$, possible for a given number of sites. We therefore define $\beta^* = \beta^*_{\mathcal{Z} =\mathcal{Z}_{\max}}$.
In the present case, we consider a lattice with $N = 14$ sites, and the corresponding maximal range is therefore, $\mathcal{Z}_{\max} = 7$. We now investigate how $\beta^*$ changes with the variation of  $\lambda$ and observe an interesting feature which we refer as \emph{rigidity}. We call the constant values obtained for $\beta^*$  with respect to $\lambda$ as rigidity. Interestingly, we observe that \(\beta^*\) obtained from the nearest neighbor entanglement shows a Hall-like plateaus with the increase of \(\lambda\) for a fixed value of \(\gamma\) (see  Figs. \ref{fig:plateau} (a) - (d)). We enumerate the observations below:
\begin{enumerate}
\item For all values of $\gamma$, the highest critical temperatures (lowest $\beta^*$ values), and maximal rigidity are obtained near $\lambda = 0$.

\item Comparing Figs. \ref{fig:plateau}(a)-(b) with Figs. \ref{fig:plateau}(c)-(d), we find that lower gamma values offer enhanced robustness, i.e., lower $\beta^*$ values in the entire $\lambda$ range inside the factorization points.

\item For $\gamma = 0.8$, we observe lower rigidity of $\beta^*$ values in comparison to other considered $\gamma$s. It can be argued by counting the number of plateaus which is found to be almost double  for covering the $\lambda$-range inside the factorization points for high values of \(\gamma\) than that of the low values of \(\gamma\).  
\end{enumerate}

In this respect,    we also analyze the minimum \(\beta\) required for obtaining  \(W\) in Eq. (\ref{eq:witness})  to be negative, when \(|\lambda| <\lambda_f^{e (p)}\) . Interestingly,  we find that although witness operator is a global characteristics of the system,  it mimics the rigidity feature obtained via  entanglement  in Fig. \ref{fig:plateau}. 

In summary, in the finite-temperature setting, we observe a wide range of novel  and counter-intuitive features in presence of variable range interactions which are not present in the nearest neighbor models. Most prominently, we report range-dependent robustness,  rigidity of robustness, and the effect of zero temperature physics on finite temperatures. 

\section{Conclusion}
\label{sec:conclusion}
Varying the range of interactions  leads to novel features in  the distribution of entanglement between different sites  in quantum spin systems. We explored these properties using a variable range anisotropic quantum $XY$ model, for which we considered the relative interaction strengths between subsequent spins to fall-off -- (1) exponentially as well as (2) polynomially (power-law decay). 

In the zero-temperature limit, on increasing the interaction range,  we ``expectedly" observed activation of several long-ranged entanglements. However, surprisingly, the activation of entanglement is not generic, and is dictated by the pair of factorization points in which the zero-temperature state is found to be a product.  
In particular, the factorization points split the  parameter-space into two disjoint regions possessing different entanglement activation rates, providing 
signatures of \emph{entanglement phase transition.}
We quantitatively confirmed these observations by computing entanglement lengths for varied interaction ranges and system parameters. 
Furthermore, we also tracked the reallocation of factorization points in the parameter-space due to the tuning in  the range of interactions. Our investigations further revealed that the distribution of entanglements follow a monogamous nature, thereby  helping us to explain features of the entanglement profile as well as entanglement length in this model.

We also analyzed the finite temperature regime in which the system suffers from thermal noise. We observed   increased robustness of entanglement with the temperature when the  model Hamiltonians involve  long-ranged interactions and are confined  between the factorization points. Specifically, we found
a hierarchy among the nearest neighbor entanglements with respect to  the range of interactions -- entanglements in canonical equilibrium states obtained from the long-ranged models  remain non-vanishing even in presence of  higher temperatures in comparison to the models involving relatively shorter range interactions which we attempted to explain using an energy-based entanglement witness. Interestingly, such an advantageous situation is present only when the system parameters lie between the factorization points.

Our work provides a systematic survey towards the control of the system parameters and interaction ranges to extract the maximal possible resource in terms of entanglement out of the zero- and finite-temperature states of the quantum spin models. We believe that  investigations in these directions can shed light on the implementation of various quantum information-theoretic protocols in quantum networks in which the distribution of entanglement plays a key role. 
\section*{Acknowledgement}
We acknowledge the support from Interdisciplinary Cyber Physical Systems (ICPS) program of the Department of Science and Technology (DST), India, 
Grant No.: DST/ICPS/QuST/Theme 1/2019/23. The authors acknowledge computations performed at the cluster computing facility of Harish-Chandra Research Institute, Allahabad, India. Some numerical results have been obtained using the Quantum Information and Computation library (QIClib).  
This research was supported in part by the ‘INFOSYS scholarship for senior students.’
\include{supple}

%
%
%
\end{document}

%% file: supple.tex
\widetext
%

%
%
%
\begin{center}
\textbf{\Large Supplementary Material \\ \Large Distribution of entanglement  with variable  range  interactions} \\
\normalsize Leela Ganesh Chandra Lakkaraju, Srijon Ghosh, Saptarshi Roy, Aditi Sen (De)\\
\small Harish-Chandra Research Institute, HBNI, Chhatnag Road, Jhunsi, Allahabad 211019, India\\
\end{center}

\setcounter{equation}{0}
\setcounter{figure}{0}
\setcounter{table}{0}
\setcounter{section}{0}
\setcounter{page}{1}
\makeatletter
\renewcommand{\theequation}{SEQ\arabic{equation}}
\renewcommand{\thefigure}{SF\arabic{figure}}
\renewcommand{\thetable}{ST\arabic{table}}
\renewcommand{\bibnumfmt}[1]{[SR#1]}
\renewcommand{\citenumfont}[1]{SR#1}
\renewcommand{\thesection}{SSEC\arabic{section}}
\hypersetup{pageanchor=false}
\subsection{Measure of Entanglement}
\label{sec:ent}
We intend to investigate how entanglement $(E)$ is distributed between two arbitrary sites. For this, we construct the reduced two party density matrix between sites $i$ and $j$ after tracing out all the parties of a $N$-party state, $\varrho_N$, except $i$ and $j$, given by  
\begin{align}
\varrho_{ij} = \text{tr}_{\overline{ij}} ~\varrho_N,
\label{eq:rhoij}
\end{align}
where $\overline{ij}$ denotes all the spins except $i$ and $j$. $\varrho_N$ is either the zero-temperature state or the canonical equilibrium state of the model, consisting of \(N\) spins. Owing to the translational invariance,
all reduced density matrices with $|i-j| = r$ of this model are identical. Therefore, $\varrho_{ij}$ only depends on the distance between the spins $i$ and $j$, and  without loss of generality, we call $\varrho_{ij}$, with $|i-j| = r$ as $\varrho_{r}$. 
Numerical analysis reveals that for the entire range of the system parameters of this model, $\varrho_r$ for all $r$ is an ``$X$"-state. It is called $X$-state  since   only non-zero elements in the density matrix $\varrho$ are the diagonal elements $\varrho_{ii}$s,  $\varrho_{23}$ and $\varrho_{14}$. Hence the density matrix  can be divided into two blocks and entanglement can be easily calculated although  the exact coefficients depend on the system parameters and $r$.

We quantify the entanglement content of $\varrho_{r}$ using logarithmic negativity  (LN) \cite{neg4, neg5}. This measure comes out of the partial transposition criterion \cite{peres} which gives a necessary and sufficient condition of entanglement for two-qubits \cite{necessary}. For an arbitrary bipartite state $\varrho_{AB}$, logarithmic negativity $E$ can be computed as
\begin{equation}
E(\varrho_{AB}) = \log_2 [2 {\cal N}(\varrho_{AB}) + 1].
\label{eq:LN}
\end{equation}
Here $\mathcal{N}$ is the negativity \cite{neg4} of $\varrho_{AB}$, defined by
\begin{equation}
  {\cal N}(\varrho_{AB})=\frac{\|\varrho_{AB}^{T_A}\|_1-1}{2},
  \label{eq:negativity}
 \end{equation}
where $\|\varrho\|_1 \equiv \mbox{tr}\sqrt{\varrho^\dag \varrho}$ is the trace-norm of the  matrix $\varrho$, with $T_A$ being the partial transposition with respect to party $A$. Throughout the manuscript, we use the logarithmic negativity to measure entanglement of $\varrho_r$, and  $E(\varrho_r) \equiv E_r$. 
Since, $\varrho_r$ is an $X$-state, its entanglement can be expressed in a closed form in terms of the state parameters. For a general $X$-state, the logarithmic negativity can  be given in a closed form in terms of the state parameters as
\begin{align}
E(\varrho) = \log_2 \big[- 2 \min \lbrace \lambda_1, \lambda_2, 0 \rbrace + 1 \big], 
\end{align}
where,
\begin{equation}
\lambda_{1 (2)} =\frac{1}{2} \times
\varrho_{11 (33)} + \varrho_{22 (44)} - \sqrt{(\varrho_{11 (33)} + \varrho_{22 (44)})^2 - 4|\varrho_{23 (14)}|^2}.\\
\end{equation}
We will use this formula to compute the entanglement of the $\varrho_r$s.


\subsection{Entanglement Length}

We are interested to determine the trends in the spread of entanglement, $E_r$ with the introduction of variable range interaction. Specifically, we want to find out, $r$, the distance upto which  $\varrho_r$ remains entangled. 
If we can rewrite $E_r$ as
\begin{align}
E_r = a + be^{-\frac{r}{\xi}}, 
\label{eq:length}
\end{align}
we call $\xi$ as the entanglement length where $a$, and $b$ are constants which can be determined from the entanglement behavior for a specific Hamiltonian. 
In  a subsequent section, we investigate the improvement of $\xi$ obtained for entanglement due to variable range interactions  in different parameter regimes.

\subsection{Factorization Points}

The quantum XY model possess a unique pair of points in the parameter-space of $\lambda$ for which the zero-temperature states of the $XY$ model are completely separable, thereby unentangled. It is a counter-intuitive feature from the perspective of entanglement resource theory, since it is not adiabatically connected to any other factorized state of the model. 
These points, for a given range of interaction $\mathcal{Z}$, can be denoted by $\lambda_{f}^{e(p)}(\mathcal{Z})$, and are called the factorization points \cite{Sfac-ger, Sfacpt}, where the superscripts $e(p)$ indicate the exponential (power-law) fall-off features.  At these points, the zero-temperature state is of the form $|\psi \rangle_N = \prod_{i=1}^N |\psi_{i} \rangle $. 
Note that for the nearest neighbor model ($\mathcal{Z}=1$), in the thermodynamic limit, $\lambda^{e(p)}_f(1)=\pm \sqrt{1-\gamma^2}$ \cite{facpt}. 
Furthermore, for the special case $(\mathcal{Z}=1)$, the quantum $XY$ Hamiltonian remains same for both exponential and polynomial fall-off scenario. Therefore, in this case, one may omit the $e (p)$ labels, and thus henceforth, we call
 $\lambda^{e(p)}_f(1)$ as $\lambda_f(1)$.

\begin{figure}[h] 
\centering
\includegraphics[width = 0.6\linewidth]{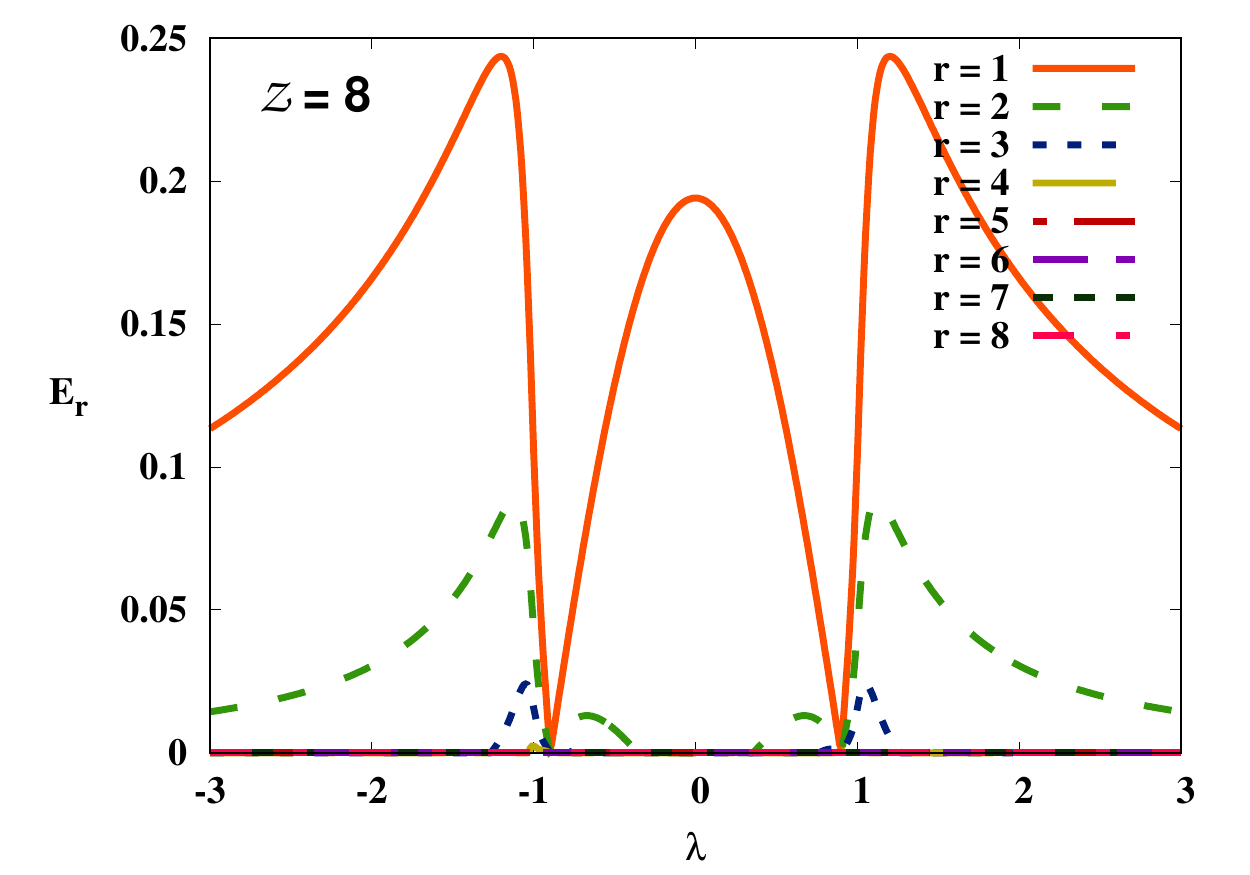}
\caption{ (Color online.)  Set of bipartite entanglements,  $\{E_r\}$ (vertical axis) with $r = 1 \ldots 8$ against  $\lambda$ (horizontal axis). Here $\mathcal{Z} = 8$ and \(\alpha_p = 5\). Here $N = 16$ and $\gamma = 0.5$. Both the axes are dimensionless.}
\label{fig:highalpha}
\end{figure}

 \subsection{Dependencies on anisotropy and  fall-off rates}
 
In this section, we explore the effects of $\gamma$ and $\alpha$ on  entanglement profiles with the introduction of variable range interactions. 
\begin{figure*}[ht] 
\centering
\includegraphics[width=0.99\linewidth]{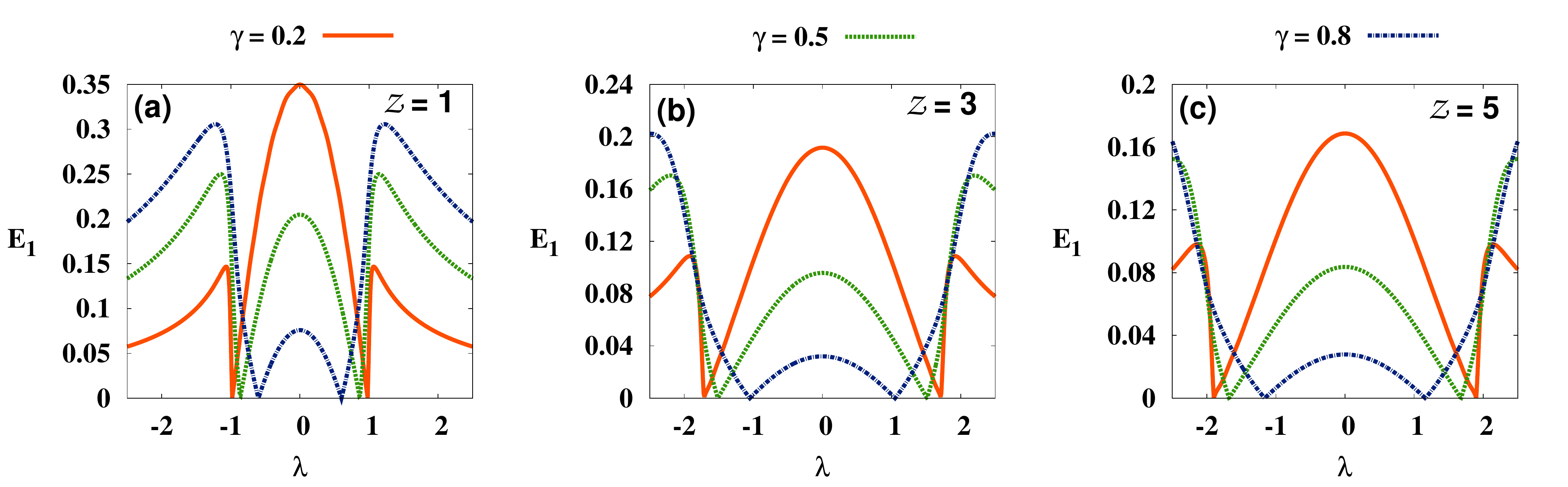}
\caption{(Color online.) Effects of anisotropy on entanglement.  \(E_1\) (vertical axis) is plotted against \(\lambda\) (horizontal axis) for three different values of \(\gamma\). Here \(\gamma =0.2, 0.5, 0.8\). (a)-(c): Different relative interaction strengths, namely \(\mathcal{Z} =1, 3, 5\) respectively. Here $N=16$ with exponential fall-off, $\alpha_e = 2$.  Both the axes are dimensionless.}
\label{fig:anisotropy}
\end{figure*}

\begin{figure*}[ht] 
\centering
\includegraphics[width=0.99\linewidth]{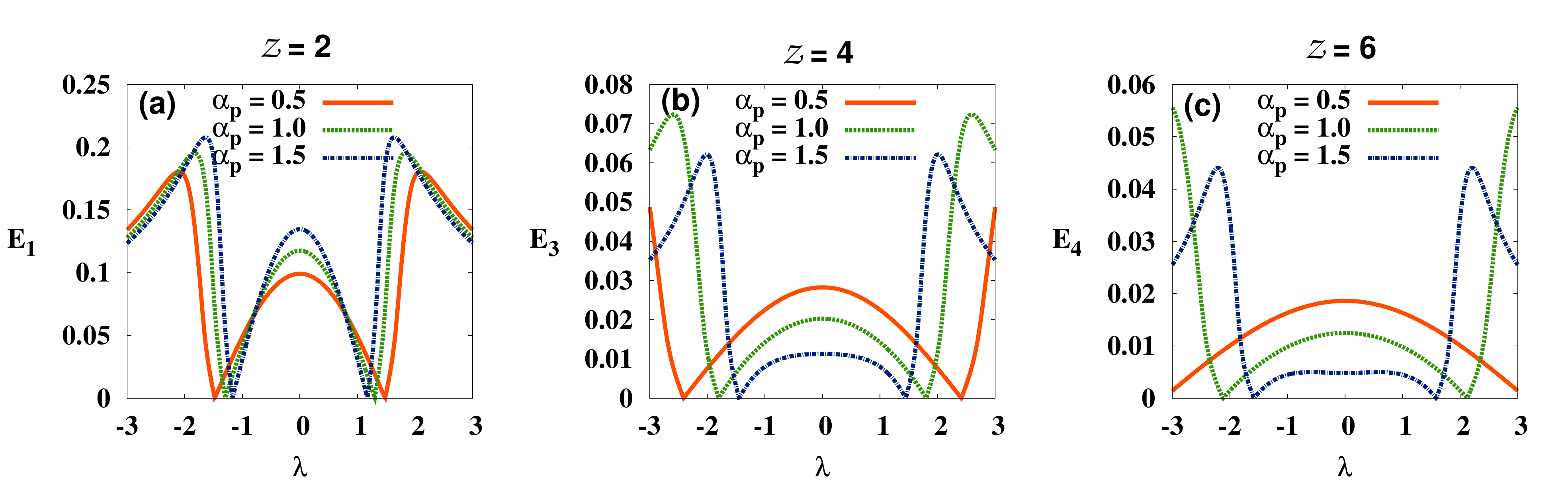}
\caption{(Color online.) Dependencies on power-law fall-off rate, $\alpha_p$. (a)-(c):  Entanglements, $E_1$, $E_3$ and $E_4$ (vertical axis) are plotted with respect to \(\lambda\) (horizontal axis) for $\mathcal{Z} = 2, ~4$, and $6$ respectively. The considered rates $(\alpha_p = 0.5, ~1, ~\text{and} ~1.5)$ are for the power-law decay of the relative interaction strengths. Here $N = 16$ and $\gamma = 0.5$.  Both the axes are dimensionless.}
\label{fig:alpha}
\end{figure*}
\emph{Dependence on anisotropy.}
 We observe that for low  values of \(\gamma\) (see $\gamma = 0.2$ in each Figs. \ref{fig:anisotropy} (a)-(c) ), i.e., when we approach  the $XX$-model, the gap between the factorization points increases, and  the $E_1$-hump between the factorization points grows. Consequently, $E_1^{\text{max}}$ comes from within the factorization points. As $\gamma$ increases further, the maximal entanglement values inside and outside the factorization points become comparable as depicted in Figs. \ref{fig:anisotropy} (a) -(c)  for \(\gamma=0.5\). 
 When \(\gamma \rightarrow 1\)  i.e., in the Ising limit, the factorization 
points come closer and the $E_1$-hump within the factorization points flattens by decreasing its magnitude in comparison to $E_1$ values obtained outside the factorization points. Therefore, for large values of $\gamma$, $E_1^{\text{max}}$ is obtained beyond the factorization points as shown in Figs. \ref{fig:anisotropy} (a)-(c) for \(\gamma =0.8\) with different range of interactions. 
Note that although the above observations are presented when the relative interaction strengths follow an exponential decay with $\alpha_e = 2$, the qualitative feature remains same even for the power-law decay.

\emph{Effects of fall-off rates on entanglement.} As in the main article, it has been discussed that the characteristics of entanglement remains almost same for both types of fall-offs (exponential and power-law). Hence, all the observations presented below hold for both the fall-offs and we skip the subscripts of \(\alpha\) . 
Note that for a given interaction range $\mathcal{Z}$, the coupling strength between distant spins (spaced not more than $\mathcal{Z}$ sites apart) are comparatively larger for lower values of $\alpha$.
Therefore, for a given $\mathcal{Z}$, a lower $\alpha$ indicates a slower decay of the relative interaction strengths which  is expected intuitively to
facilitate the enhancement (if not activation) of long-ranged entanglements, see Fig. \ref{fig:alpha} (b) and (c). 
 Therefore, it is tempting to take the above argument one step forward and  expect that lower values of $\alpha$ are always ``better" than that of the high values of \(\alpha\) with respect to entanglement enhancement or activation. 
This intuition holds in almost all cases involving longer-ranged interactions.  However, we find that for any given interaction range, $\mathcal{Z}$, the nearest neighbor entanglement, $E_1$, possess a higher values for higher  $\alpha$s. From Fig. \ref{fig:alpha} (a),  we notice that for $\mathcal{Z} = 2$, \(\alpha_p= 1.5\) leads to high nearest entanglement content, $E_1$   compared to the  case with $\alpha_p = 0.5$. 
Such a behavior can again be explained in the light of a monogamy-based argument. In particular, 
 for a given interaction range,  enhancement or activation of longer-ranged entanglements comes at the expense of reducing  the shorter ranged ones owing to constraints set by the monogamy relation.
Therefore, for a fixed $\mathcal{Z}$,  one has to resort to  lower $\alpha$ values  to create a large amount of long-ranged entanglements while it is wiser to choose high values of $\alpha$ for maximizing short-ranged entanglements. This feature of entanglement is independent of the choices of anisotropy parameters, \(\gamma\).

\emph{Effects of moderate $\alpha_p$.} When $\alpha_p$ is moderate, the bipartite entanglement other than nearest neighbor entanglement is very small. In Fig. \ref{fig:highalpha},  we consider, $\alpha = 5$ and $N = 16$ while the Hamiltonian is fully connected, i.e., with $\mathcal{Z} = 8$. 
 Specifically, when $|\lambda| <  \lambda_f$,   $E_r, r > 1$ is almost negligible while when the $|\lambda|>\lambda_f$, we were able to see non-zero entanglement, for $r = 1,2 \text{ and } 3$ with significant values. Precisely we notice that with moderate value of $\alpha$, the system behaves almost as a nearest neighbor model. The trend of \(\{E_r\}\)  shows that For a fixed \(\mathcal{Z}\),  one can distinguish whether \(\alpha_p\) is small or \(\alpha_p\) is moderately high.

%



%% file: vr02Nov_arxiv_replacement.bbl
\begin{thebibliography}{58}
\bibitem [{mx()}]{mx}%
  \BibitemOpen
  \bibinfo {note} {The magnetization along the $x$-direction is defined by
  \begin{eqnarray} m_x = \frac{1}{N}\sum_i^N m_x^i = \frac{1}{N} \sum_i^N
  \text{tr}(\sigma_x^i \varrho_i), \nonumber \end{eqnarray} where $\sigma_x^i$
  represents Pauli $x$ matrix, and $\varrho_i$ is the single site reduced
  density matrix of $i^{\text{th}}$ site of the ground state of the Hamiltonian
  in Eq. \eqref{eq:ham1} perturbed by an infinitesimal field along the
  $x$-direction. For translationally invariant models (the case considered
  here), $m_x^i$ becomes site independent, and can be written as
  \begin{eqnarray} m_x = \text{tr}(\sigma_x^1 \varrho_1) \nonumber
  \end{eqnarray}}\BibitemShut {NoStop}%
\end{thebibliography}
